\documentclass[12pt]{article}
\pdfoutput=1
\usepackage{jheppub}
\usepackage{amsmath}
\usepackage{amsfonts}
\usepackage{amssymb}
\usepackage{graphicx}

\usepackage[export]{adjustbox}
\newcommand{\bmat}{\left(\begin{array}}
\newcommand{\emat}{\end{array}\right)}

\def\yzero{\smash{\hbox{$y\kern-4pt\raise1pt\hbox{${}^\circ$}$}}}

\def\beq{\begin{equation}}
\def\eeq{\end{equation}}
\def\beqa{\begin{eqnarray}}
\def\eeqa{\end{eqnarray}}

\def\-{\hphantom{-}}

\def\s2{\frac{1}{\sqrt2}}

\def\beq{\begin{equation}}
\def\eeq{\end{equation}}
\def\beqa{\begin{eqnarray}}
\def\eeqa{\end{eqnarray}}

\def\IF{\relax{\rm I\kern-.18em F}}
\def\II{\relax{\rm I\kern-.18em I}}

\def\Dsl{\,\raise.15ex\hbox{/}\mkern-13.5mu D} 

\def\IC{{\bf C}}
\def\IS{{\bf S}}

\def\IX{{\bf X}}

\def\IP{\bf P}

\def\NN{{\cal N}}

\catcode`\@=11   
\newdimen\@rotdimen
\newbox\@rotbox  

\def\@vspec#1{\special{ps:#1}}
\def\@rotstart#1{\@vspec{gsave currentpoint currentpoint translate
   #1 neg exch neg exch translate}}
\def\@rotfinish{\@vspec{currentpoint grestore moveto}}
\def\@rotr#1{\@rotdimen=\ht#1\advance\@rotdimen by\dp#1%
   \hbox to\@rotdimen{\hskip\ht#1\vbox to\wd#1{\@rotstart{90 rotate}%
   \box#1\vss}\hss}\@rotfinish}
\def\@rotl#1{\@rotdimen=\ht#1\advance\@rotdimen by\dp#1%
   \hbox to\@rotdimen{\vbox to\wd#1{\vskip\wd#1\@rotstart{270 rotate}%
   \box#1\vss}\hss}\@rotfinish}%
\def\@rotu#1{\@rotdimen=\ht#1\advance\@rotdimen by\dp#1%
   \hbox to\wd#1{\hskip\wd#1\vbox to\@rotdimen{\vskip\@rotdimen
   \@rotstart{-1 dup scale}\box#1\vss}\hss}\@rotfinish}%
\def\@rotf#1{\hbox to\wd#1{\hskip\wd#1\@rotstart{-1 1 scale}%
   \box#1\hss}\@rotfinish}%
\def\rotate{\@ifnextchar[{\@rotate}{\@rotate[l]}}
\def\@rotate[#1]#2{\setbox\@rotbox=\hbox{#2}\@nameuse{@rot#1}\@rotbox}

\catcode`\@=12

\begin{document}

\makeatletter
\@addtoreset{equation}{section}
\makeatother
\renewcommand{\theequation}{\thesection.\arabic{equation}}
\pagestyle{empty}
\rightline{ IFT-UAM/CSIC-18-121}
\vspace{1.2cm}
\begin{center}
\LARGE{\bf Transplanckian Axion Monodromy !?
}
\large{Ginevra Buratti, Jos\'e Calder\'on,  Angel M. Uranga\\[4mm]}
\footnotesize{Instituto de F\'{\i}sica Te\'orica IFT-UAM/CSIC,\\[-0.3em] 
C/ Nicol\'as Cabrera 13-15, 
Campus de Cantoblanco, 28049 Madrid, Spain}

\vspace*{5mm}

\small{\bf Abstract} \\
\end{center}
\begin{center}
\begin{minipage}[h]{17.0cm}
We show that warped throats of the Klebanov-Strassler kind, regarded as 5d flux compactifications on Sasaki-Einstein manifolds $\IX_5$, describe fully backreacted solutions of transplanckian axion monodromy. We show that the asymptotic Klebanov-Tseytlin solution features a 5d axion physically rolling through its dependence on an spatial coordinate, and traversing arbitrarily large distances in field space. The solution includes the backreaction on the breathing mode of the compactification space and on the vacuum energy, which yields a novel form of flattening. We establish the description of the system in terms of an effective 5d theory for the axion, and verify its validity in transplanckian regimes. In this context, rolling axion monodromy configurations with limited field space range would correspond, in the holographic dual field theory, to duality walls, which admit no embedding in string theory so far. We present an identical realization of transplanckian axion monodromy in 4d in fluxed version of AdS$_4\times \IX_7$.
We speculate that similar models in which the axion rolls in the time direction naturally correspond to embedding the same mechanism in de Sitter vacua, thus providing a natural arena for large field inflation, and potentially linking the swampland de Sitter and distance conjectures.
\end{minipage}
\end{center}
\newpage
\setcounter{page}{1}
\pagestyle{plain}
\renewcommand{\thefootnote}{\arabic{footnote}}
\setcounter{footnote}{0}

\tableofcontents

\vspace*{1cm}

\section{Introduction and conclusions}

The application of Quantum Gravitational constraints to string theory vacua has motivated several conjectures limiting the possibilities to attain field ranges larger than the Planck scale \cite{Vafa:2005ui,Ooguri:2006in,ArkaniHamed:2006dz,Ooguri:2016pdq,Obied:2018sgi,Harlow:2018jwu,Harlow:2018tng} (see \cite{Brennan:2017rbf} for a recent review). A prototypical case is the application of the Weak Gravity Conjecture \cite{ArkaniHamed:2006dz} to axions, which implies that periodic axion potentials, such as those in natural inflation, cannot host transplanckian field ranges \cite{Montero:2015ofa}. Another general result is the Swampland Distance Conjecture, which states that as some modulus approaches a point at infinite distance in moduli space, there is an infinite tower of states becoming massless, exponentially with the distance \cite{Ooguri:2006in}. There are also partial studies concerning axion monodromy models \cite{Silverstein:2008sg}, see also \cite{McAllister:2008hb,Flauger:2009ab,Arends:2014qca,Marchesano:2014mla,McAllister:2014mpa,Franco:2014hsa,Ibanez:2014kia,Retolaza:2015sta}, trying to rule out their transplanckian excursion by invoking the backreaction on the scalar kinetic terms reducing the effectively traversed distance \cite{Baume:2016psm} \footnote{For other discussions of backreaction related to flattening of the potential, see \cite{Dong:2010in}.}. There are also discussions ruling out particular models using 10d lifts \cite{McAllister:2016vzi, Kim:2018vgz} or other mechanisms \cite{Blumenhagen:2014nba}.

These results would seem to motivate a Swampland Transplanckian conjecture, stating that transplanckian field ranges are not physically attainable in Quantum Gravity. If correct, this statement would have profound implications for certain phenomenological applications, like the construction of inflation models with sizable gravitational wave backgrounds (which for single-field inflation are directly related to the distance traversed by the inflaton). The purpose of the present article is to prove that this conjecture is in fact incorrect, and that transplanckian field excursions are physically realized in string theory. We will do it by presenting a completely explicit example of axion monodromy model, with full backreaction taken into account, in terms of the complete 10d supergravity solution. The complete background turns out to be given by a simple and well-known warped throat, the Klebanov-Strassler throat \cite{Klebanov:2000nc,Klebanov:2000hb}, when regarded as a flux compactification on a Sasaki-Einstein manifold $\IX_5$, with a 5d axion rolling in the radial direction of a (locally) AdS$_5$ spacetime.

Let us make some relevant remarks:

$\bullet$ The discussion is intended as an existence proof of transplanckian axion monodromy in string theory. In particular, we focus on discussing how the 10d solution fully encodes the backreaction of the axion dynamics including the impact on axion kinetic terms, and the backreaction on other sectors, including the compactification moduli and the vacuum energy. This last point is extremely relevant and has not been taken into account in earlier attempts to describe 10d lifts of axion monodromy models.

$\bullet$ We consider configurations where the axion has an explicit dependence on the non-compact spacetimes coordinates (in fact, on a particular spatial coordinate). This is crucial for the configuration to allow the axion to climb its potential while maintaining the solution on-shell. Again, this is an ingredient not properly accounted for in earlier analysis of 10d backreaction of transplanckian axion monodromy, and ties directly to the question of including the vacuum energy variation in the analysis. 

$\bullet$ On the other hand, it is {\em physically meaningful} to consider configurations where the axion is actually varying in spacetime. After all, the main motivation for scalars with transplanckian field excursions are large field inflation models, in which the eventual cosmological solution corresponds to a time-dependent configuration of the scalar rolling down its potential. 

$\bullet$ We work in configurations with negative vacuum energy. This is not an obstruction from the fundamental viewpoint of establishing the existence of transplanckian field excursions in string theory. On the other hand, it does not yield realistic models for inflation. Related to this, our configurations have axions depending on spatial directions, rather than time-dependent ones. In fact, formally the sign flip required to switch from space to time dependent scalar profiles correlates with the sign flip for the vacuum energy. This suggests a tantalizing link between positive cosmological constant and time dependent background, which in the present context is reminiscent of the dS/CFT correspondence \cite{Strominger:2001pn}. It would be interesting to explore the relation of our transplanckian axion monodromy scenario with recent discussions of bounds on slow-roll and the swampland de Sitter conjectures \cite{Garg:2018reu}.

$\bullet$ We focus on 5d models because the kinds of Klebanov-Strassler throats we need (either for the conifold or for generalizations) have been most studied in this setup. On the other hand, there are less studied but completely analogous throats based on locally AdS$_4\times \IX_7$ configurations in M-theory, which we also discuss and lead to 4d transplanckian axion monodromy configurations in precisely the same fashion as the 5d models. 

$\bullet$ The dynamics of the transplanckian axion can be described within an effective field theory, which we discuss explicitly based on a consistent truncation provided in \cite{Klebanov:2000nc}. This, together with the full 10d solution, allows for a discussion of the validity of effective actions for the transplanckian excursion. We show that the configuration is free from oftentimes feared problems: no pathology arises neither when the axion winds its period a large number of times, and no infinite tower of states becomes exponentially light when the axion reaches beyond transplanckian distances in field space.

$\bullet$ Freund-Rubin vacua such as AdS$_5\times\IX_5$ with 5-form flux on $\IX_5$ are often described as not yielding good effective field theories, since the compactification radius is comparable to the AdS radius. However, we are not interested in describing an effective field theory which describes the stabilization of the compactification breathing mode, which cannot be decoupled (in the Wilsonian sense) from the KK tower of states. We are interested in the effective dynamics of a massless axion and its spacetime variations at much lower scales, and in its backreaction effects, which are also controlled by those scales. Our effective theory is suitable for that purpose, and can be regarded as describing the low energy dynamics of a scalar in a gravitational background which is fixed at higher scales, save for backreaction effects which are duly included in the effective field theory description.

\medskip

The paper is organized as follows. In Section \ref{sec:general-throats} we describe the KS solutions from the perspective of producing 5d axion monodromy models, focusing on the conifold example. In section \ref{sec:5d} we describe the 5d compactification on $\IX_5$ with no 3-form fluxes, leading to the AdS$_5$ vacuum. In section \ref{sec:the-throat} we describe the KS solution \cite{Klebanov:2000hb} (actually, its KT asymptotic form \cite{Klebanov:2000nc}) and in section \ref{transplanckian1} we establish that it describes an axion monodromy solution in which the field range traversed is arbitrarily large, in particular transplanckian. In section \ref{sec:duality-walls} we relate hypothetical backgrounds with finite axion field ranges with duality walls in the UV of the holographically dual field theories, which have so far not been shown to admit a gravitational description. In Section \ref{sec:eft} we turn to the effective field theory description. In section \ref{sec:kt-eft} we review the effective field theory in \cite{Klebanov:2000nc} for the axion and compactification moduli. In section \ref{sec:axion-eft} we obtain an effective action at energies hierarchically below the KK scale, which actually encodes the axion dynamics and its backreaction effects. In section \ref{sec:fourdcase} we discuss 4d configurations from M-theory compactifications, with exactly the same axion monodromy physics as the previous 5d examples. Appendix \ref{appendix} discusses a dual Hanany-Witten configuration of D4- and NS5-branes useful to illustrate the absence of pathologies as the axion winds around its period.

\medskip

\section{Warped throats and transplanckian axion monodromy}
\label{sec:general-throats}

In the following we review the Klebanov-Strassler (KS) throat \cite{Klebanov:2000hb}. We intentionally emphasize its structure as a 5d compactification in which the introduction of the RR 3-form flux yields a 5d axion monodromy model, for which the KS throat is an explicit fully backreacted solution. We then show that the axion roll in this configuration is transplanckian. Actually, for this purpose it suffices to focus on the region far from the tip of the throat, so we use the simpler expressions of the Klebanov-Tseylin (KT) throat \cite{Klebanov:2000nc}, supplemented with the boundary conditions derived from the KS smoothing of its naked singularity. For the latter reason, we still refer to the configuration as KS throat.

\subsection{The 5d theory}
\label{sec:5d}

Consider as starting point the type IIB Freund-Rubin AdS$_5\times T^{1,1}$ background
\beqa
ds^2\, =\, R^2 \,\frac{dr^2}{r^2} \, +\, \frac{r^2}{R^2} \, \eta_{\mu\nu} dx^\mu dx^\nu\, +\, R^2\, ds_{T^{1,1}}^2
\label{ads-t11}
\eeqa
with
\beqa
R^4\, =\, 4\pi (\alpha')^2 g_s N
\label{R-value}
\eeqa
and with $N$ units of RR 5-form flux through $T^{1,1}$. The type IIB complex  coupling is constant, and we will keep it set at $\tau=i/g_s$ (introduction of non-trivial constant $C_0$ is straightforward via minor changes in the fluxes below).

This is the near horizon limit of a set of $N$ D3-branes at a conifold singularity \cite{Klebanov:1998hh}. The line element $ds_{T^{1,1}}^2$ corresponds to a (unit volume) 5d horizon $T^{1,1}$, which is an $\IS^1$ bundle over $\IP_1\times \IP_1$ with first Chern classes $(1,1)$, hence the name. Topologically, it is an $\IS^2\times\IS^3$. Denoting by $\sigma_2$ and $\sigma_2'$ the volume forms of the two $\IP_1$'s, we have a harmonic 2-form $\omega_2=\sigma_2-\sigma_2'$ and its (dual in $T^{1,1})$ harmonic 3-form $\omega_3$. They are Poincar\'e duals of the 3- and 2-spheres, and  $\omega_2\wedge \omega_3$ is the volume form on $T^{1,1}$.

On top of the complex dilaton, the resulting effective 5d theory has a massless axion, given by the period of the NSNS 2-form over $\IS^2\subset T^{1,1}$
\beqa
\int_{\IS^2} B_2=\phi \quad {\rm namely}\quad B_2\,=\,\phi\,\omega_2.
\label{def-phi}
\eeqa
The periodicity  $\phi\sim \phi+1$ is set by the exponential of the action of a  fundamental string wrapped on the $\IS^2$.
Above the scale of massless fields, there is the scale $1/R$. This is the scale of KK modes, but also the scale of stabilization of the breathing mode of $T^{1,1}$. It is possible to write an effective action for this dynamical mode\footnote{Since this scale is not hierarchically lower than the KK masses, this effective action should be interpreted as arising from a consistent truncation, rather than a Wilsonian one.}; in this action, the potential is minimized at the value (\ref{R-value}), and with a negative potential energy cosmological constant, such that the maximally symmetric solution is the AdS$_5$ space in (\ref{ads-t11}). For a simplified discussion in the completely analogous case of AdS$_5\times \IS^5$, see \cite{Silverstein:2004id}; we will discuss such effective actions in a more general context later on.

The above background is a particular case of the general class of AdS$_5\times \IX_5$ vacua, where $\IX_5$ is a Sasaki-Einstein variety. These are gravitational duals to systems of D3-branes at singularities, and have been intensely explored in the literature. Large classes of these models admit also the introdution of 3-form fluxes to be described below, and thus lead to axion monodromy models. To emphasize this direct generalization, we will oftentimes write $\IX_5$ instead of $T^{1,1}$.

\subsection{The KS solution}
\label{sec:the-throat}

Once we have described the compactification to 5d, we would like to describe the introduction of a RR flux on $\IS^3\subset T^{1,1}$
\beqa
\frac{1}{(2\pi)^2 \alpha'}\int_{\IS^3} F_3\, = \, M.
\label{rr-flux}
\eeqa

Our key observation is that the resulting 5d theory is an axion monodromy model for $\phi$. This simply follows because the self-dual 5-form field strength 
\beqa
{\tilde F}_5\, =\, dC_4\, -\frac 12 \, C_2\wedge H_3\, +\, \frac 12 \, B_2 \wedge F_3
\eeqa
satisfies the modified Bianchi identity 
\beqa
d*{\tilde F}_5\, =\, d{\tilde F}_5\, =\, H_3\wedge F_3.
\label{bianchi}
\eeqa

From the KK perspective the flux (\ref{rr-flux}) induces a 5d topological coupling 
\beqa
\int_{10d} F_3\wedge B_2 \wedge F_5 \quad \longrightarrow M\int_{5d} \, \phi\, F_5.
\eeqa
As already noted in \cite{Franco:2014hsa,Buratti:2018onj} \footnote{While finishing this paper, we noticed the recent \cite{Hebecker:2018yxs}, which involves a similar structure of flux and axion, albeit in a different approach to axion monodromy.}, this is a 5d version of the Dvali-Kaloper-Sorbo term \cite{Dvali:2005an,Kaloper:2008fb} associated to a monodromy for the axion. Clearly, as $\phi$ winds around its basic period, there is a corresponding increase for the flux of ${\tilde F}_5$ through $T^{1,1}$ (and, by self-duality, through the non-compact 5d space), as follows,
\beqa
N\, =\, \int_{T^{1,1}}\, {\tilde F}_5\, =\, N_0\, +\,  M\phi .
\label{monodromic-n}
\eeqa
In the following we take the reference value $N_0$ to be reabsorbed into a redefinition of $\phi$.

The presence of a scalar potential of the axion monodromy kind, arising from the reduction of the 10d $|{\tilde F}_5|^2$ terms, will be manifest in the 5d effective action discussed in Section \ref{sec:eft}. We are interested in the behaviour of this theory as the value of $\phi$ changes over a large range. Clearly, the presence of this potential term implies that moving the scalar vev adiabatically away from the minimum leads to off-shell configurations, for which the computation of the backreaction is not clearly defined. A natural solution is to instead consider configurations in which the scalar $\phi$ is allowed to roll, so that the spacetime dependent background allows to remain on-shell\footnote{This is in fact a natural viewpoint in inflationary axion monodromy models, in which the interesting solutions correspond to physical time-dependent rolls of the scalar down its potential.}. The KS solution is precisely an explicit 10d solution of this rolling configuration in which the axion $\phi$ is allowed to roll along one of the {\em spatial} directions. (As discussed in the introduction, the realization of time dependent roll suggests an interesting interplay with the question of realizing de Sitter vacua). We now review the 10d KS solution (actually, its KT limit with KS boundary conditions) from this perspective.

The KS throat describes a configuration in which the axion has a dependence on the radial direction. Concretely, $\phi$ is a harmonic form in the radial direction in the underlying AdS$_5$, hence
\beqa
\Delta \phi = 0 \quad \to \quad \phi \sim M \log r .
\eeqa
This corresponds to the fact that the combination $G_3=F_3-\frac{i}{g_s}H_3$ is imaginary self-dual, and in fact (2,1) i.e. supersymmetry preserving, when regarded as a flux in the conifold CY threefold $\IX_6$, i.e. when combining the radial coordinate $r$ with the angular manifold $T^{1,1}$. The metric then simply corresponds to a warped version of $M_4\times \IX_6$ of the general class in \cite{Dasgupta:1999ss,Giddings:2001yu}
\beqa
ds^2_{10} =   h^{-1/2}(r)   dx_n dx_n 
 +  h^{1/2} (r)  (dr^2 + r^2 ds^2_{T^{1,1}} )
\eeqa
with 
\beqa
h(r)=   \frac{1}{4r^{4}}{ M^2 \log {\frac{r}{ r_*}} }
\label{harmonic-function-kt}
\eeqa
with $r_*$ some reference value.
In short, the metric is of the form (\ref{ads-t11}) with the radius (\ref{R-value}) including a radial dependence
\beqa
N\sim M^2\,\log r,
\label{rdep-n}
\eeqa
which follows from (\ref{monodromic-n}).
As explained, this is the KT solution, which has a naked singularity at $r\to 0$. The KS solution provides a smoothing of this based on the deformed conifold\footnote{When regarded from the 5d perspective, this implies that the direction $r$ ``ends'' at a finite distance. Of course this is not relevant for the discussion below, which only deals with the large $r$ regime. Moreover, even if one would be interested in having a radial dimension with no end, it is straightforward to modify (\ref{harmonic-function-kt}) or even its full KS version, e.g. by introducing a large number $P$ of additional explicit D3-branes, producing an AdS$_5$ at the bottom of the KS throat, effectively removing the endpoint for $r$. This corresponds to the mesonic branches of the cascade \cite{Dymarsky:2005xt}.}. In fact we will be interested in the region of large $r$, and how it extends to infinity, so the KT solution suffices.

The above solution describes precisely all the effects of the backreaction for arbitrarily large values of the axion and number of windings along its period. As one moves towards large $r$, the axion is climbing up its potential and inducing larger flux $N$ due to the monodromy. The flux and stored energy backreact on the stabilization of the breathing mode of the compactification space, whose minimum tracks the value of $\phi$ from (\ref{R-value}), (\ref{monodromic-n}) and
 (\ref{rdep-n})
\beqa
R^4\,\sim \, g_s M\phi\, \sim\, g_s M^2 \log r.
\eeqa
The non-compact geometry is locally  AdS$_5$ with varying radius $R$. Hence, there is also a backreaction in the vacuum energy, with runs towards less negative values as
\beqa
V_0\sim (\log r)^{-1}.
\eeqa
The slow growth of the vacuum energy can be regarded as a flattening of the potential, albeit different from the polynomial ones in \cite{Dong:2010in}.

\medskip

From the holographic perspective, each winding of $\phi$ on its period corresponds to a cycle in the cascade of Seiberg dualities, in which, as one moves to the UV (larger $r$), the effective number of colors increases by (actually twice) a factor $M$
\beqa
&& SU(N_0)\times SU(N_0+M) \,\rightarrow\, SU(N_0+2M)\times SU(N_0+M) \,\rightarrow \nonumber \\
&&\quad\quad  \,\rightarrow\, SU(N_0+2M)\times SU(N_0+3M).
\eeqa
Although we will not exploit this holographic picture (as the supergravity solution speaks for itself), we will use it in Appendix \ref{appendix} to explain why no disaster arises when the axion rolls around its period\footnote{See \cite{Flauger:2009ab} for some discussion of periodic effects in axion monodromy.}. In particular there are no states becoming massless or light as one crosses the ``zero'' value, an effect often feared to play a lethal role for the discussion of monodromy dynamics in effective field theory. The fact that this effect is absent in our model supports the expectation that it is not a generic problem of axion monodromy models (but rather, either of particular models realizing the idea, or of partial analysis of those models without full inclusion of backreaction).

\subsection{Transplanckian axion field range}
\label{transplanckian1}

Let us use the above solution to quickly show that the 5d field $\phi$ traverses a transplanckian distance in field space. A more systematic discussion is presented in Section \ref{sec:eft}.

The distance traversed by $\phi$ from a reference point $r_0$ to infinity is given by
\beqa
\Delta\, =\, \int_{r_0}^\infty\, \left(\, G_{\phi\phi} \, \frac{d\phi}{dr}\, \frac{d\phi}{dr}\, \right)^{\frac 12}\, dr\, =\, 
 \int_{r_0}^\infty\, \left(\, G_{\phi\phi} \right)^{\frac 12}\, \frac{d\phi}{dr}\,  dr,
 \label{distance-gen}
\eeqa
where $G_{\phi\phi}$ is the metric in field space, which is determined by the 5d kinetic term for $\phi$, in the 5d Einstein frame
\beqa
 S_{ 5} \, = \,  \frac{1}{2\kappa_{5}^2}\,   \int d^{5} x \, \sqrt{-g_{5}}\, 
 \bigg( \,  {\cal R}_5  \, - \,  G_{\phi\phi}\, \partial_m \phi\, \partial_n \phi\, g^{mn}\, \bigg).
\eeqa
Since the compactification volume varies, certain care is required. We must define a fixed reference radius $R$ determining the 5d Planck scale, and introduce a 5d dynamical breathing mode ${\tilde R}$ encoding any variation (see \cite{Silverstein:2004id} for a similar parametrization). Hence, focusing just on the parametric dependence, we write
\beqa
V_{\IX_5}\, =\, R^5\, {\tilde R}^5,
\eeqa
\beqa
ds^2\, =\, g^{(5)}_{mn}\, dx^m\, dx^n\, +\, (R{\tilde R})^2 \, (g_{\IX_5})_{ij}\, dy^i\, dy^j.
\eeqa
We now focus on the reduction on $\IX_5$ of the 10d action for the metric and kinetic term of $B_2$. In the 10d Einstein frame we have
\beqa
S_{10d}\, =\, \frac{1}{2\kappa_{10}^{\,2}}\, \int d^{10} x \, \sqrt{-g_{10}}\, 
 \bigg( \,  {\cal R}_{10}  \, - \,  \frac{1}{12\, g_s}\, H_{MNP}\, H^{MNP}\, \bigg).
\eeqa
As explained, the reference value $R$ fixes the 5d Planck scale
\beqa
\frac{R^5}{2\kappa_{10}^{\,2}}= \frac{1}{2\kappa_{5}^2}
\eeqa
and the factor ${\tilde R}^5$ is reabsorbed by rescaling the 5d metric to the 5d Einstein frame
\beqa
(g_5)_{mn}\, \to \, {\tilde R}^{-\frac{10}{3}}\, (g_5)_{mn}.
\eeqa
We follow the effect of this rescaling in the kinetic term of the component of $B_2$ given by (\ref{def-phi}). The dependence on ${\tilde R}$ is as follows:
\beqa
&&   \int\, d^{10}x\, \sqrt{-g_{10}}\, g^{mn}g^{ik}g^{jl}\partial_{m} B_{ij} \partial_{n}B_{kl}\,\stackrel{\rm compact.}{\longrightarrow}\\
 &&\stackrel{\rm compact.}{\longrightarrow}\, \int\, d^5x\, \sqrt{-g_{5}} \, (R{\tilde R}) \, (g_{5})^{mn} \, \partial_{m}\phi\, \partial_n\phi\, \stackrel{\rm Einstein}{\longrightarrow} \, \int\, d^5x\, \sqrt{-g_{5}} \, (R{\tilde R}^{-4}) \, (g_{5})^{mn} \, \partial_{m}\phi\, \partial_n\phi\,\nonumber.
 \label{kinetic-term-5d-string-frame}
\eeqa
Hence, we have ${\tilde R}^4\sim M^2 \log r$ and thus
\beqa
G_{\phi\phi}\sim (\, M^2\, \log r\,)^{-1}.
\label{kinetic-coeff1}
\eeqa
We have $\phi\sim M\log r$, hence the distance (\ref{distance-gen}) is
\beqa
\Delta\,=\, \int \, G_{\phi\phi}^{\frac 12}\,  \frac{d\phi}{dr}\, dr\, \sim\,  \int \,dr \,(M^2\, \log r)^{-\frac 12} \, M\, \frac {dr}r\, = \, \int \frac{ds}{s^{\frac 12}}
\eeqa
for $s=\log r$. This becomes arbitrarily large for large $r$, showing that the 5d scalar $\phi$ rolls through a transplanckian distance in field space.

The 10d backreacted solution for this transplanckian axion monodromy configuration allows to address  many of the objections to transplanckian field excursions in string theory or quantum gravity, and study how the present models avoid those potential pitfalls. As many of these are related to the regimes of validity of effective field theories for the axion dynamics, we postpone their discussion until section \ref{sec:eft}.

\medskip

The above AdS$_5$ vacua admit generalizations associated to D3-branes at more general CY threefold singularities, which have been extensively studied in the toric case. The dual backgrounds correspond to type IIB Freund-Rubin AdS$_5\times \IX_5$, where $\IX_5$ is the 5d horizon of the 6d CY cone. The construction of KT backgrounds by introducing (possibly a richer set of) 3-form fluxes is a straightforward extension of our above discussion (see for instance \cite{Franco:2004jz} for complex cones over del Pezzo surfaces), so there is a large class of constructions leading to transplanckian axion monodromy. Being more careful, we should make clear that only CY singularities admitting complex deformations can complete their KT throats into smooth supersymmetric KS-like throats \cite{Franco:2005fd}; other choices admit no supersymmetric KS completion \cite{Berenstein:2005xa,Franco:2004jz,Bertolini:2005di}, and actually lead to runaway instabilities \cite{Franco:2004jz,Intriligator:2005aw}, a fact which has recently motivated the ``local AdS - Weak Gravity Conjecture" \cite{Buratti:2018onj}, generalizing the ``AdS-WGC" in \cite{Ooguri:2016pdq}. However, even with the restriction to CY singularities admitting complex deformations, there is an enormous class of such explicit constructions (built with standard toolkits, see e.g. \cite{GarciaEtxebarria:2006aq}), and thus leading to transplanckian axion monodromy.

\subsection{Duality walls}
\label{sec:duality-walls}

The fact that the axion traverses an arbitrarily large distance in field space as one moves to larger distances in $r$ is intimately related to the RG flow structure in the holographic field theory. As mentioned in section \ref{sec:the-throat}, the axion winding around its period corresponds to completing a cycle in the Seiberg duality cascade of the $SU(N)\times SU(N+M)$ field theory. The steps in the energy scale in each duality cycle relate to the radial distance required for the scalar to wind around its period. The infinite range in energy as one moves up to the UV in the field theory provides an infinite range in radial distance on the gravity side, which allows for an arbitrarily large axion field range with finite gradient energy density. Hence, the nice properties of the holographic field theory RG flow relates to the fact that the gravity side is described by a supergravity background. 

In contrast with this picture, it is interesting to point out that a different kind of RG flow behaviour of duality cascades has been contemplated, purely from the field theory perspective. These are known as duality walls, and correspond to duality cascade RG flows in which, as one moves to the UV, the energy steps in each duality cycle decrease; more concretely, the number of duality cycles in a given energy slice increases as one moves up to the UV, in such a way that there is a limiting energy, at which the number of cycles per energy interval diverges. Such RG flows have been introduced in \cite{strassler-wall}, and proposed to relate to quiver gauge theories of D-branes at singularities in e.g. \cite{Hanany:2003xh,Franco:2003ja,Franco:2003ea}. However, there is no concrete string theory D-brane realization of such RG flows. In particular, systematic searches for gravity backgrounds dual to gauge theories with duality walls have produced no such results \cite{Franco:2004jz}. 

The absence of such backgrounds, at least in the context of supergravity, has an interesting implication for our perspective on field ranges in axion monodromy models. Gravitational solutions dual to duality walls would require an axion winding around its period an infinite number of times in a finite range in the radial distance. This is compatible with finite gradient energy densities only if the kinetic term of the axion varies so as to render finite the traversed distance in field space. This kind of behaviour would produce axion monodromy models where superplanckian field ranges cannot be attained. Hence, the absence of supergravity backgrounds of this kind is a signal that superplanckian axion monodromy models are actually generic in the present setup, whereas those with limiting field ranges are exotic, if at all existent.

\section{Effective field theory analysis}
\label{sec:eft}

In the previous section we have shown a fully backreacted explicit 10d solution for axion monodromy models with arbitrarily large field ranges. In this section we bring the discussion to the context of the 5d effective field theory, where much of the discussion of swampland conjectures is carried out.

\subsection{Effective field theory for axion and breathing mode}
\label{sec:kt-eft}

From the 10d solution it is clear that the relevant dynamics in 5d involves the axion $\phi$ and the breathing mode of $\IX_5=T^{1,1}$, coupled to 5d gravity. It is interesting to device an effective field theory describing the dynamics for these degrees of freedom in the KS solution \footnote{Inclusion of the dilaton is discussed in section \ref{sec:non-geodesic}}. This provides a concrete context in which to test the regime of validity of the effective field theory to describe transplanckian axion monodromy, or to test other swampland conjectures.

The 5d effective field theory can be obtained starting from the 10d type IIB effective action, and using a suitable ansatz for the compactification, which allows for general dynamics for the relevant 5d fields. This strategy was in fact put forward in \cite{Klebanov:2000nc} to produce the 5d action we are interested in. We review the key ingredients relevant for our purposes, and adapted to our present notation.

We consider the metric ansatz
\beqa
&&ds^2_{10} = L^2 \big( \, e^{- 5q}\, ds^2_5 + \, e^{3q}ds^2_{T^{1,1}} \big)\ .  
\label{metric-ansatz-eft}
 \eeqa
Here $q$ is a 5d field encoding the breathing mode of $T^{1,1}$.  Also, $ds^2_5$ is the line element in the 5d non-compact spacetime, defined in the 5d Einstein frame thanks to the prefactor $e^{-5q}$. The explicit $L$ scales out the line elements to geometries of unit radius. 

There are $M$ units of $F_3$ flux over the $\IS^3\in T^{1,1}$ and there is a 5d axion defined by (\ref{def-phi}). The modified Bianchi identity (\ref{bianchi}) implies that the flux of ${\tilde F}_5$ over $T^{1,1}$ is given by (\ref{monodromic-n}).

The 5d effective action for the 5d scalars $\phi$ and $q$, collectively denoted by $\varphi^a$, is given by 
\beqa
 S_{ 5} 
= - \frac{2}{\kappa_{5}^2}  \int d^{5} x 
\ \sqrt{-g_{5}} \,\bigg[ 
{ \frac 1 4}  R_5  
- { \frac 1  2} G_{ab}(\varphi)  \partial \varphi^a \partial \varphi^b 
- V(\varphi)\bigg],
\label{ktaction}
\eeqa
with the kinetic terms and potential given by 
\beqa
&& G_{ab}(\varphi)  \partial \varphi^a \partial \varphi^b \, =\,  15 (\partial q)^2    + { \frac{1}{4}}  g_s^{-1}e^{ - 6 q}(\partial \phi )^2 \ , \label{kinetic}\\
&& V(\varphi) \, = \,-5 e^{ - 8 q} + {1\over 8} M^2 g_s\, e^{- 14 q}
 + {1\over 8} (N_0 +  M \phi)^2 e^{ - 20 q}. \label{potential}
\eeqa
The  different terms in the potential have a clear interpretation. The first negative contribution corresponds to the curvature of the compactification space $T^{1,1}$, the second is the contribution from the $M$ units of $F_3$ flux on the $\IS^3$, and the third corresponds to the contribution from the 5-form flux over $T^{1,1}$, and has the typical axion monodromy structure. We note that, despite the bare quadratic dependence, the backreaction of $\phi$ on the geometry will produce a different functional dependence of the potential energy at the minimum, as shown below. Also, as already explained, the above action should be regarded as a consistent truncation in supergravity, so we will take special care to discuss the role of other physical degrees of freedom, like KK modes.

Since the above effective theory is general, it should reproduce the basic AdS$_5$ background for $M=0$. The potential becomes
\beqa
V(\varphi) \, = \,-5 e^{ - 8 q} \, +\, {1\over 8} N_0^{\,2} e^{ - 20 q} .
\label{ads-potential}
\eeqa
The potential has a minimum at
\beqa
e^{6q}\, =\, \frac{N_0}{4}
\eeqa
with negative potential energy at the minimum
\beqa
V_0\, =\, -3\, e^{-8q}.
\label{v0}
\eeqa
Comparing (\ref{metric-ansatz-eft}) with the standard expression for AdS$_5\times T^{1,1}$ metric (\ref{ads-t11}), we recover the scaling of the $T^{1,1}$ radius $R$ with $N_0$
\beqa
R^2\, \sim\, e^{3q} \, \rightarrow \, R^4\sim N_0   
 \eeqa
with other factors reabsorbed in $L$ in (\ref{metric-ansatz-eft}). Taking the value for $V_0$ (\ref{v0}) and removing a factor of $e^{-5q}$ to change to the 10d frame, we recover the same scaling for the radius of the AdS$_5$ vacuum. 

\medskip

The KS throat (actually its asymptotic KT form) is a solution of the above effective action. Following \cite{Klebanov:2000nc}, we take the following ansatz for the metric
\beqa
ds^2_{10} \,=\,   s^{-1/2}(r) \,  \eta_{\mu\nu}\, dx^\mu\, dx^\nu\, +\,  h^{1/2} (r)\,  (dr^2 + r^2 ds^2_{T^{1,1}} ) .
\eeqa
In terms of (\ref{metric-ansatz-eft}), this corresponds to 
\beqa
 e^{3 q }  = r ^2 \, h^{1/2} (r)\quad , \quad
ds_5^{\, 2}\, =\, e^{5q} \, [\,s^{-1/2}(r) \, \eta_{\mu\nu}\, dx^\mu\, dx^\nu\, +\,  h^{1/2} (r)  \,dr^2 \,].
\label{relation}
\eeqa

The effective theory admits a solution where
\beqa
\phi\, =\, M\, \log r\quad , \quad s(r)\, =\, h(r)\, =\,   \frac{1}{4r^{4}}{ M^2 \log {\frac{r}{ r_*}} },
\eeqa
with $r_*$ some reference value. This is just the throat solution discussed in Section \ref{sec:the-throat}.

\medskip

The effective action can be exploited to recover the result of the transplanckian field range covered by the axion. Since the 
 5d effective action is already in the 5d Einstein frame, we can read out and evaluate the kinetic term for $\phi$ in (\ref{kinetic})
\beqa
G_{\phi\phi}\, \sim\, e^{-6q}\, =\, [\,r^4\, h(r)\,]^{-1} \, \sim\,  (\, M^2\, \log r\, )^{-1}.
\eeqa
We thus recover, in a more precise setting, the result (\ref{kinetic-coeff1}), and thus the corresponding unbounded (and hence transplanckian) field range.

\subsection{The axion effective field theory}
\label{sec:axion-eft}

As explained, the above action should be regarded as a consistent truncation in supergravity, but not as a Wilsonian effective action. In other words, at the scale $1/R$ at which the stabilization of the breathing mode occurs, there are many other modes, corresponding to KK excitations of the 10d fields in $\IX_5$ which are not included in the action. Note that this scale goes as $1/R\sim (\log r)^{-1/4}$. On the other hand, the effective dynamics for the axion occurs at far lower scales, set by $\partial \phi=1/r$. Similarly, the scale of the backreaction on the compactification radius or the vacuum energy is measured by their derivatives with respect to $r$, which are similarly suppresed by $1/r$ (or even with additional inverse powers of $\log r$). It is therefore interesting to construct an effective field theory including just the axion and intended to describe its dynamics at those scales (hence, including the backreaction on the volume and vacuum energy).

For this, we minimize the scalar potential for $q$ keeping $\phi$ fixed. This gives the condition
\beqa
\frac 52 \, (N_0+M\phi)^2 \, x^2\, +\, \frac 74\,  g_s M^2\, x \, -\, 40\,=\, 0\quad, \quad {\rm with}\; x=e^{-6q}.
\label{second-deg}
\eeqa
Rather than solving the above exactly, since we are focusing on the large $r$ regime, where $\phi$ is large and $x$ is comparably small, we drop the subleading second term, and obtain
\beqa
e^{6q}\, =\,\frac 14 (N_0\,+\, M\phi) .
\label{track}
\eeqa
This reproduces the result of the KS solution that $e^{6q}\sim M^2\log r$ for $\phi\sim M\log r$, so we are capturing the relevant physics. 

We should replace that value in the potential. Again restricting to large $r$, we drop the second term in (\ref{potential}) and obtain
\beqa
V\, =\, - e^{-8q}\, \big[\,  5\, -\, \frac 18 \, (N_0\, +\, M\phi)^2\, e^{-12q}\,\big].
\eeqa
This has the same structure as (\ref{ads-potential}) with the replacement $N_0\to N_0+M\phi$. The potential should be regarded as a function of $\phi$ only, by simply replacing (\ref{track}) in this expression. 
It is therefore clear that considering a profile $\phi=M\log r$ leads to the appropriate change in the vacuum energy, so that the backreaction of the axion monodromy is duly included. 

The complete axion action should include its kinetic term, obtained from that in (\ref{kinetic}) by using (\ref{track}). We recover a kinetic term
\beqa
\sim (N_0+M\phi)^{-1} (\partial \phi)^2,
\eeqa
which again reproduces the familiar result about the transplanckian distance traveled in the rolling solution considered.

\medskip

This effective action suffices to describe the dynamics of the transplanckian axion monodromy, so it is a well-defined setup to test/propose swampland conjectures on effective actions. For instance, one natural idea is to consider if there is an analog of the swampland distance conjecture, and there is a tower of states becoming exponentially light as the axion travels at arbitrarily large distances. This is not the case, as follows. The invariant distance in axion field space goes (for large $\phi$) as $d\sim \phi^{1/2}$; on the other hand, the masses of KK modes (which are the primary suspects for fields becoming light at large $\phi$, since $R$ increases), scale as $m_{\rm KK}\sim e^{-4q}\sim (\phi)^{-2/3}$, hence $m_{\rm KK}\sim d^{-4/3}$ and there is no tower of exponentially light states. This is compatible with the swampland distance conjecture, if interpreted as applying to field ranges approaching points at infinite distance in moduli space \cite{Ooguri:2006in,Grimm:2018ohb}. It is also compatible with the oftentimes used version for transplanckian geodesic distances, since in the next section we will show that our axion travel does not follow a geodesic. However the model provides a beautiful way in which a fully backreacted monodromic axion can travel arbitrarily large distance  in field space without triggering the appearance of exponentially light states.

There are other interesting questions that can be addressed in the present setup, such as the application of swampland constraints on the scalar potential, or the realization of the weak gravity conjecture in the present setup, etc. Since the underlying model is a string theory compactification on a smooth geometry with fluxes, we expect no new surprises or novel mechanisms related to these other swampland conjectures. 

\subsection{Inclusion of the dilaton}
\label{sec:non-geodesic}

As announced, in this section we show that the underlying reason for the compatibility of the transplanckian axion monodromy model with the swampland distance conjectures is that the axion does not follow a geodesic in the moduli of light fields. The crucial ingredients to understand this are the spacetime dependence of the axion, and the inclusion of the dilaton in the moduli space.

The original KT 5d effective action \cite{Klebanov:2000nc} includes further fields beyond those included in the earlier discussion. Indeed, it contains fields $\varphi^a=q,f,\Phi,\phi$, where $f$ describes a possible asymmetric volume for the ${\bf S}^2$ and ${\bf S}^3$ of $T^{1,1}$, and $\Phi$ is the dilaton. The 5d action for these fields has the structure (\ref{ktaction}) with
\beq\begin{split}
G_{ab}(\varphi)&=\operatorname{diag}\left(15,10,\frac{1}{4},\frac{1}{4}e^{-\Phi-4f-6q}\right),\\
V(\varphi)&=e^{-8q}\left(e^{-12f}-6e^{-2f}\right)+\frac{1}{8}M^2e^{\Phi+4f-14q}+\frac{1}{8}(N_0+M\phi)^2e^{-20q}.
\end{split}\eeq
The pure AdS$\times T^{1,1}$ solution for $M=0$ shows that in this action the breathing mode $q$ and asymmetric mode $f$ are heavy modes, while the axion $\phi$ and dilaton $\Phi$ remain as light fields. Morally, we should thus consider the later as parametrizing a moduli space at scales hierarchycally below the KK scale, with a potential induced by the introduction of non-zero $M$. This is manifest because the terms including $M$ in the potential are subdominant with respect to the first, $M$-independent, one.

This allows to integrate out $q$ and $f$. We may minimize the leading potential for $f$, and set $f=0$ (as implicit in the previous section). For the minimization of $q$, we proceed as in the previous section and recover (\ref{track}).

Note that, in the resulting theory for the axion and the dilaton, there is a non-trivial potential for the dilaton. This is however compatible with its constant value in the axion monodromy solution in an interesting way:  the spacetime dependence of the axion has a non-trivial backreaction in the dilaton, through the dilaton dependence of the axion kinetic term, which induces an effective potential for the dilaton balancing the original one and allowing for a constant dilaton solution. Quantitatively, the equation of motion for a general field in the presence of a spacetime-dependent axion background reads
\beq
\label{eomSCALARS}
\frac{1}{\sqrt{g}}\partial_\nu\left(\sqrt{g}g^{\mu\nu}G_{ac}\partial_\mu\varphi^a\right)=\frac{1}{2}\frac{\partial{G_{\phi\phi}}}{\partial{\varphi^c}}(\partial \phi)^2+\frac{\partial{V}}{\partial{\varphi^c}}.
\eeq
For the dilaton, the condition to allow for a constant dilaton $e^{\Phi}=g_s$ is the vanishing of the right-hand side, which is proportional to
\beqa
-e^{-6q-\Phi}(\partial\phi)^2+e^{-14q+\Phi}M^2 .
\eeqa
This indeed vanishes in the KT solution, allowing for a constant dilaton. As anticipated, the spacetime dependence of the axion exerts a force on the dilaton keeping it constant on the slope of its bare potential.

The scale of this effect is set by the gradient of the axion $\partial \phi$, which is hierarchycally below the KK scale. This implies that the corresponding backreaction effect for the other fields $q$ and $f$ is negligible, and can be ignored when they are integrated out, as implicit in our above discussion. It also implies that it is not appropriate, in a Wilsonian sense, to integrate out the dilaton dynamics, as it occurs at the scale relevant for axion dynamics.

This last observation raises an important point. In checking the interplay of our axion monodromy model with the swampland distance conjectures, the moduli space on which distances should be discussed is that spanned by the axion and the dilaton, as their potential on this moduli space is hierarchically below the KK scale cutoff. As we have shown, in this moduli space the KT solulion describes an axion monodromy model traversing transplanckian (and actually arbitrarily long) distances without encountering infinite towers of light states. However, as we now argue, this does not  contradicts swampland distance conjectures, since the trajectory does not correspond to a geodesic in the axion-dilaton moduli space.

After replacement of $q$ and $f$ by their values at the minimum of their potentials, the kinetic term for $\phi$, $\Phi$ reads
\begin{equation}
\mathcal{L}_{kin}=\frac{1}{8}(\partial\Phi)^{2}+\left(\frac{e^{-\Phi}}{2\left(N_{0}+M\phi\right)}+\frac{5M^{2}}{24\left(N_{0}+M\phi\right)^{2}}\right)(\partial\phi)^{2}.
\end{equation}
At large $\phi$ we can neglect the subleading second term in the kinetric term of $\phi$ and get
\begin{equation}
\mathcal{L}_{kin}=\frac{1}{8}(\partial\Phi)^{2}+\frac{e^{-\Phi}}{2\left(N_{0}+M\phi\right)}(\partial\phi)^{2}.
\end{equation}
To look at the geodesics of this theory it is convenient to change variables
\begin{equation}
\begin{matrix}x=\frac{4}{M}\sqrt{N_{0}+M\phi},\\
y=2e^{\Phi/2}=2\sqrt{g_{s}}.
\end{matrix}
\end{equation}
This leads to 
\begin{equation}
\mathcal{L}_{kin}=\frac{1}{2y^{2}}\left[(\partial x)^{2}+(\partial y)^{2}\right],
\end{equation}
which is the metric of the hyperbolic plane. Geodesics of this space, considering $y$ the vertical
axis, are vertical lines or half-circles centered in the horizontal axis. On the other hand, the KT solution corresponds to horizontal
lines at different constant values of the dilaton.

\section{The 4d case}
\label{sec:fourdcase}

The above discussion has been carried out in the 5d context because, being holographically dual to 4d gauge theories, these are the best studied warped throats. However, there are well studied supergravity solutions of the form AdS$_4\times \IX_7$, and supergravity solutions of the KT kind when the horizon variety $\IX_7$ admits the introduction of fluxes \cite{Herzog:2000rz}.
In the following we review these backgrounds and show that they realize in 4d the same kind of transplanckian axion monodromy as the 5d configurations described above.

The starting point is the AdS$_4\times \IX_7$ background, which can be regarded as arising from the near-horizon limit of a stack of $N$ coincident M2-branes \cite{Maldacena:1997re}
\beqa
ds^2\, =\, h(r)^{\frac 23}\, \eta_{\mu\nu} \, dx^\mu\, dx^\nu\, +\, h(r)^{\frac 13}\, (\, dr^2\, +\, r^2\, ds_{\IX_7}^{\, 2}\, ),
\label{metric-kh}
\eeqa
where now Greek indices label non-compact coordinates spanning, together with $r$, the 4d spacetime.
The harmonic function is
\beqa
h(r)\, =\, \frac{2^5\, \pi^2\, N\, \ell_p^{6}}{r^6}.
\eeqa
Namely, we have
\beqa
ds^2\, =\, \frac{R^4}{r^4}\, \eta_{\mu\nu} \, dx^\mu\, dx^\nu\, +\, R^2 \,\frac{dr^2}{r^2}\, +\, R^2\, ds_{\IX_7}^{\, 2}\, ,
\eeqa
where
\beqa
R^6\, =\, 2^5\, \pi^2\, N\, \ell_p^{6} .
\eeqa
There are $N$ units of flux of the 7-form field strength $F_7$ (dual to the 4-form field strength $F_4$) through $\IX_7$.

Consider an $\IX_7$ with a non-trivial 4-cycle\footnote{Such horizons can be obtained for instance by taking the near horizon limit of M2-branes at toric CY$_3\times \IC$ (leading to 3d $\NN=1$ theories), where the CY$_3$ admits a complex deformation corresponding to the size of a 3-cycle. The horizon $\IX_7$ then contains (an $\IS^1$ worth of) such 3-cycle, and hence its dual 4-cycle.}, on which we turn on $M$ units of 4-form field strength flux $F_4$. Taking the dual 3-cycle $\Pi_3$ in $\IX_7$, there is a 4d axion 
\beqa
\phi\, =\, \int_{\Pi_3} C_3.
\eeqa
This axion is monodromic, as follows from the reduction of the 11d Chern-Simons coupling
\beqa
\int_{11d} F_4\wedge F_4\wedge C_3\, \rightarrow\, \int_{4d} \, M\, \phi\, F_4 .
\eeqa
The monodromy implies that the value of $N$ varies with $\phi$ as
\beqa
N\, =\, N_0\, +\, M\,\phi ,
\eeqa
with $N_0$ a reference value, which we take zero in what follows.

This leads to a 4d analog of the KT throat found in \cite{Herzog:2000rz} and given by a flux background
\beqa
F_4\,=\, d^3x\wedge dh^{-1}\, +\, M\, *_7\omega_3\, -\, M\, \frac{dr}{r}\wedge \omega_3 .
\eeqa
Here $\omega_3$ is the Poincare dual to the 4-cycle in $\IX_7$, so the second term corresponds to the $F_4$ flux through the 4-cycle. The third term corresponds to a rolling scalar profile $d\phi=dr/r$, hence
\beqa
\phi\, \sim\, M\log r .
\eeqa
Hence we have the axion rolling  logarithmically up its monodromic potential, exactly as in the the 5d KS solutions discussed above.
The first term correspond to the dual of the flux of $F_7$ through $\IX_7$, which varies with the radial coordinate due to the axion monodromy. 

The harmonic function $h(r)$ is
\beqa
h(r)\, =\, M^2\, \big(\, \frac{\log r}{6r^6} \, +\, \frac{1}{36r^6}\, \big)
\eeqa
(up to some $\rho/r^6$ factor, which defines a reference value which we take to be zero). It also determines the metric by replacement in (\ref{metric-kh}). 

The solution, just like in the 5d KT example, has a naked singularity at $r=0$, which is presumably smoothed out at least for certain geometries $\IX_7$, although no analog of the full KS solution has been found. It would be interesting to develop the dictionary of fractional M2-brane theories and their gravity duals further to gain insight into such smoothings. This however lies beyond the scope of the present paper.

\medskip

It is straightforward to compute the 4d kinetic term of the axion $\phi$ as in the simplified 5d calculation in section \ref{transplanckian1}. Specifically, the  Einstein-Hilbert and 3-form kinetic term in the 11d action read
\beqa
S_{11}\, =\, \frac{1}{2\kappa_{11}^{\,2}}\, \int\, d^{11}x\, \sqrt{-g_{11}}\, \big(\, {\cal R}_{11}\, +\, \frac 12 \, |F_4|^2\, \big) .
\eeqa
Define the volume of $\IX_7=(R{\tilde R})^7$, where $R$ defines the backgound value and ${\tilde R}$ its breathing mode. The KK reduction to 4d contains the terms
\beqa
S_{4}\, =\, \frac{1}{2\kappa_{4}^{\,2}}\, \int\, d^{11}x\,  \sqrt{-g_{4}}\, \big(\, {\tilde R}^7\, {\cal R}_4\, +\, c\,  {\tilde R}\, g^{mn}\, \partial_m \partial_n\phi \, \big) .
\eeqa
Here we have introduced 
\beqa
\kappa_4^{\, 2}\, =\, \frac{\kappa_{11}^{\,2}}{R^7} .
\eeqa
Also, the factor ${\tilde R}$ in the axion kinetic term arises from an ${\tilde R}^7$ from the compactification volume and a factor ${\tilde R}^{-6}$ from three inverse metrics of $\IX_7$ required for the contractions of $|\omega_3|^3$. Finally $c$ is a constant that depends on geometrical properties of the cycles in $\IX_7$. 

Going to the 4d Einstein frame we have
\beqa
S_{4}\, =\, \frac{1}{2\kappa_{4}^{\,2}}\, \int\, d^{11}x\,  \sqrt{-g_{4}}\, \big(\, {\cal R}_4\, +\, c\,  {\tilde R}^{-6}\, g^{mn}\, \partial_m \partial_n\phi \, \big) .
\eeqa
So the kinetic term for the axion gives
\beqa
G_{\phi\phi}\, \sim\, {\tilde R}^{-6}\, \sim\, (M^2\log r)^{-1} .
\eeqa
This is exactly as in the 5d example, and again leads to arbitrarily large, in particular transplanckian, field ranges traversed by the axion roll.

\section*{Acknowledgments}
We are pleased to thank S. Franco, L. Ib\'anez, F. Marchesano, M. Montero, C. Vafa  and I. Valenzuela  for useful discussions. This work is partially supported by the grants  FPA2015-65480-P from the MINECO/FEDER, the ERC Advanced Grant SPLE under contract ERC-2012-ADG-20120216-320421 and the grant SEV-2016-0597 of the ``Centro de Excelencia Severo Ochoa" Programme. The work by J.C. is supported by a FPU position from Spanish Ministry of Education.

\newpage

\appendix

\section{Periodic crossing and the dual Hanany-Witten picture}
\label{appendix}

In this section we discuss a T-dual realization of the KS duality cascade, in terms of the NS5- and D4-brane configurations \cite{Witten:1997sc} realizing 4d gauge theories \`a la Hanany-Witten \cite{Hanany1997}. The picture is similar to that mentioned in \cite{Silverstein:2008sg}, albeit with additional relevant refinements.

The configuration is flat 10d space with one dimension, labelled 6, compactified on an $\IS^1$. There is one NS5-brane along the directions 012345 (and at the origin in 89), and one NS5-brane (denoted NS5') along the directions 012389 (and at the origin in 45), with D4-branes along 0123 and suspended among them in 6 (and at the origin in 4589), in a compact version of \cite{Elitzur:1997hc}. The positions of all branes in the directions 7 are taken equal. The numbers of D4-branes at each side of the interval are $N$ and $N+M$ respectively. The scalar $\phi$ corresponds to the distance (in units of $2\pi$ the radius of $\IS^1$) between the NS and the NS'-branes, so it has periodicity $\phi\sim \phi + 1$. 

In a naive description, as the scalar winds around its period, the crossings of the NS and NS'-branes produce Seiberg dualities that complete a full cycle in the duality cascade. This naive picture would seem to suggest that each crossing leads to additional light degrees of freedom, which could spoil the axion monodromy, or at least its description in terms of an effective action not including these new modes.

However, the actual picture is somewhat more intricate and is free of these problems. The answer lies in the phenomenon of brane bending in \cite{Witten:1997sc}, which implies that the $M$ additional D4-branes on one of the intervals forces the NS- and NS'-branes to bend. This bending has a logarithmic dependence, and is a long distance result of the description of the whole system as a single M5-brane in a holomorphic curve in the M-theory lift of the configuration \cite{Witten:1997sc,Witten:1997ep}. In $\NN=2$ 4d theories, this corresponds in a precise manner to the field theory running of gauge couplings on the Coulomb branch. In the present $\NN=1$ setup, the RG direction (to become the radius in the gravitational dual side) can be thought of as the radial distance away from the point $x^4=x^5=x^8=x^9=0$ at which all branes are located. Then, there is a logarithmic bending of the positions of the NS- and NS'-branes in the directions 6, which matches the above naive description. However, the other positions of the NS- and NS'-branes in the other directions do not coincide, hence no actual crossing of branes occurs. The discussion of Seiberg dualities carries over but in this more precise sense. The phenomenon is similar to the discussion in \cite{Evslin:2001sw}.

\bibliographystyle{JHEP}
\bibliography{mybib}

\end{document}